\documentclass[12pt]{iopart}

\usepackage{epsfig,graphics,graphicx}
\begin{document}

\title[Photons at RHIC: The role of viscosity and of initial state fluctuations]{Photons at RHIC: The role of viscosity and of initial state fluctuations}

\author{Maxime Dion$\,^a$, Charles Gale$\,^a$\footnote{Presenter.}, Sangyong Jeon$\,^a$, Jean-Fran\c{c}ois Paquet$\,^a$, Bj\"orn Schenke$\,^b$, Clint Young$\,^a$}

\address{$^a\,$Department of Physics, McGill University, 3600 University Street, Montreal QC, Canada H3A 2T8}
\address{$^b\,$Physics Department,  Bldg. 510A, Brookhaven National Laboratory,  Upton, NY 11973, USA}
\ead{gale@physics.mcgill.ca}
\begin{abstract}
We study real photons produced in heavy ion collisions at RHIC, and we calculate their spectrum and its azimuthal momentum  anisotropy.  The photons from a variety of sources are included, and the interplay and the time-evolution of those sources are modelled in a full 3D hydrodynamic simulation. We quantify the $v_2$ of thermal photons produced in ideal and viscous fluids, and the consequences of using different initial conditions are explored. 
\end{abstract}


\section{Introduction}
Real and virtual photons are penetrating probes, and thus their measurement can reveal much of the internal dynamics of relativistic heavy ion collisions. A complication, however, is that photons are emitted throughout the space-time history of the nuclear collisions. In addition, several sources of electromagnetic radiation exist, and these may be either signal or background, depending on the physics being pursued. Therefore, a complete picture  needs to address all these known sources, and to consider their evolution in a realistic scenario of the nuclear collision. We concentrate here on the real photons measured in heavy ion collisions at RHIC: these will include pQCD prompt photons, thermal photons from the quark-gluon plasma (QGP) and from the hadron gas (HG), and photons from jets interacting with the hot and dense medium. We calculate  photon momentum spectra and their  azimuthal anisotropy, and consider the emission from media in and out of equilibrium.
\section{Hydrodynamical evolution and results}
The nucleus-nucleus collision is modelled in this work by using \textsc{music} \cite{Schenke:2010nt}: a three-dimensional, hydrodynamic simulation of relativistic nuclear collisions. Importantly, \textsc{music} can be currently modified to include shear viscosity, $\eta$, and fluctuating initial conditions (FIC). 

Quite generally, the hydrodynamic evolution traces a space-time locus which satisfies the conservation laws for the energy-momentum tensor and the baryon current: $\partial_\mu T^{\mu \nu} = 0$, and $\partial_\mu J_{\rm B}^\mu = 0$. In the ideal case, $T_{\rm ideal}^{\mu \nu} = \left( \epsilon + P \right) u^\mu u^\nu - P g^{\mu \nu}$ and $J_{\rm B,\,  ideal}^\mu = n_{\rm B} u^\mu$. Note that $P$ is  pressure, $\epsilon$ is the local energy density, and $n_{\rm B}$ is the local baryon density; these quantities are related through the equation of state. Also, $u^\mu = \left(\gamma, \gamma {\bf v} \right)$ is the local flow velocity relative to some fixed reference frame. In cases where the dynamics are dissipative, the conservation laws are modified. More specifically,  $T^{\mu \nu} (\eta)= T_{\rm ideal}^{\mu \nu} + \pi^{\mu \nu} (\eta)$. These details will not be discussed here, but can be found in Refs. \cite{Schenke:2010rr, Schenke:2011qd,Jeon:2011}. However, it is important to state that the viscous corrections will affect not only the bulk evolution  but also the microscopic distribution functions, $f_0$. A popular ansatz \cite{Dusling:2009df} for this modification is  $f_0 \to f_0 + \delta f$, where $\delta f = f_0 
\left(1 \pm f_0\right) p^\mu p^\nu \pi_{\mu \nu} / 2 \left(\epsilon + P\right)T^2$ ($\pm$ for bosons/fermions). 
For processes where production yields are amenable to a rate equation formulation, the importance of viscous corrections are thus established by using the out-of-equilibrium distribution functions above.  It is important to realize that the viscous calculations demand a complete revaluation of all the underlying photon rates: the details will appear elsewhere \cite{Dion:2011}.
\begin{figure}[h]
\begin{center}
{\includegraphics[width=7cm]{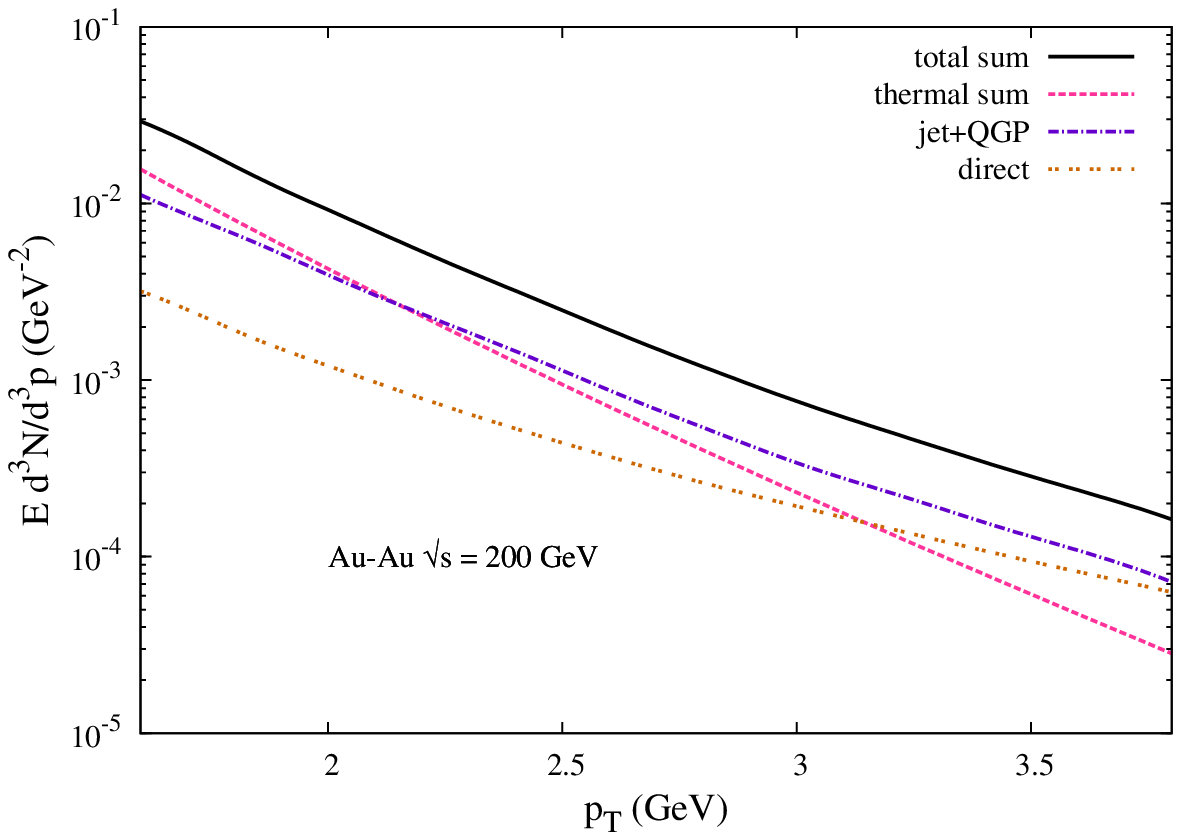}\hspace*{1cm}
\includegraphics[width=7cm]{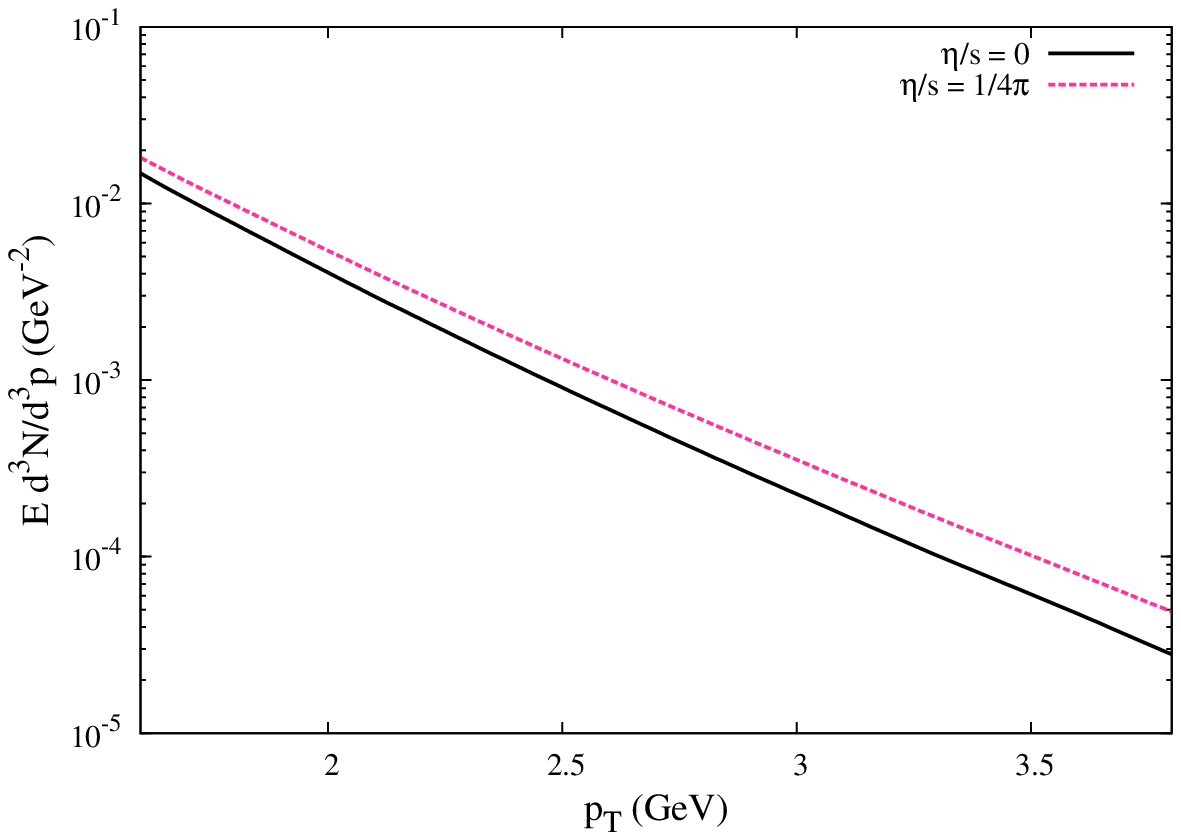}}
\caption{\sl Left panel: The yield of photons produced in Au-Au collisions at RHIC energy, in the 0-20\% centrality class. 
Shown separately are the pQCD direct photons from the primordial nucleon-nucleon collisions, calculated at NLO \cite{Paquet:2011} (all scales set to $p_T/2$), the photons from the jet-plasma interaction \cite{Turbide:2007mi,Qin:2009bk}, and the ``thermal photons'' discussed in the text. Right panel: The net thermal photon spectrum, with and without the viscous corrections (here $\eta/s = 1/ 4 \pi$, and $s$ is the entropy density). }
\label{photon_yield}
\end{center}
\end{figure}
The net yield of photons produced in relativistic heavy ion collisions is then obtained by integrating the rates for photon production  with a time-evolution approach. Finally and importantly, this model has to be constrained by data, hadronic and electromagnetic. In this context, \textsc{music} has been shown to reproduce the measured characteristics of hadronic spectra. The photon rates for equilibrium hadronic ensembles exist for the QGP  \cite{Arnold:2001ms} and the HG  \cite{Turbide:2003si} phases. The real photons generated by an ideal hydrodynamic evolution are considered first, and the results are shown on the left panel of Figure \ref{photon_yield}. To set the scale, only a few photon channels are shown and the fragmentation component of primordial nucleon-nucleon collisions has been assumed suppressed \cite{Arleo:2011}. The right panel quantifies the influence of viscous effects on the net thermal photon spectrum. Even for intermediate values of the photon transverse momentum, the changes are non-negligible. 
\begin{figure}[h]
\begin{center}
{\includegraphics[width=7cm]{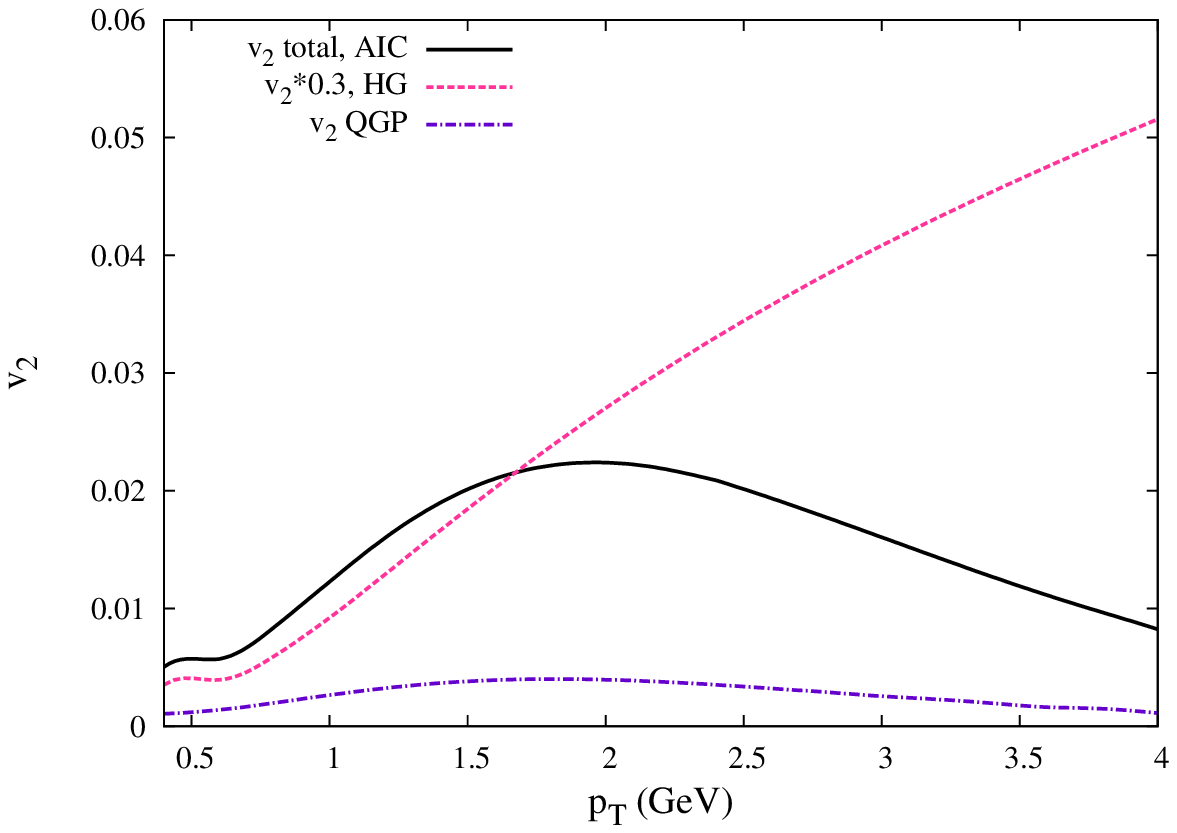}\hspace*{1cm}
\includegraphics[width=7cm]{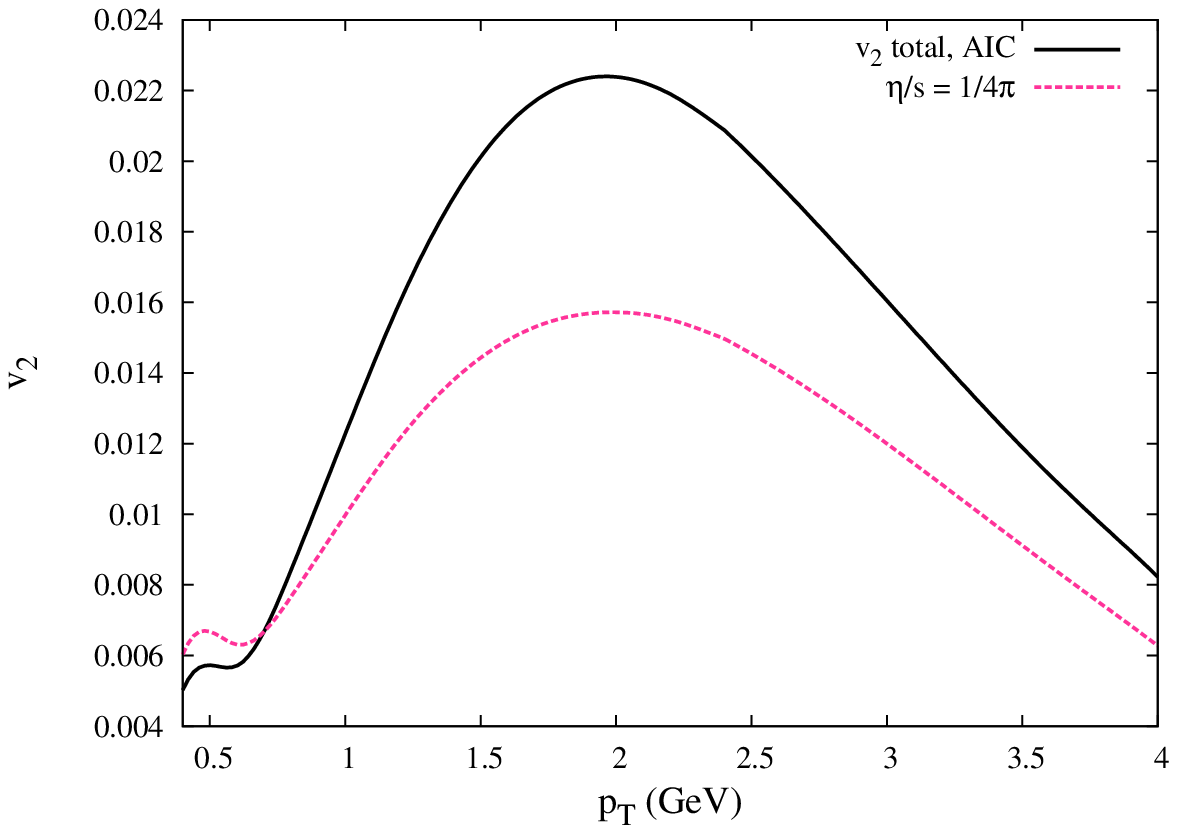}}
\caption{\sl Left panel: The $v_2$ coefficient for thermal photons is computed as a function of transverse momentum, after an ideal hydrodynamical evolution. The separate contributions are the $v_2$ from the photons emitted in the QGP phase, in the HG phase, and the net $v_2$. Right panel: The effect of viscous corrections on the net thermal photon $v_2$ is shown.}
\label{photon_v2}
\end{center}
\end{figure}
Going beyond simple one-body spectra, it is useful to state that the azimuthal anisotropy of the photon momentum distribution has been shown to be a probe sensitive to the nature of the underlying degrees of freedom \cite{Chatterjee:2005de,Heinz:2006qy}. This fact is made clear in Figure \ref{photon_v2}, where the left panel shows the value of the $v_2$ coefficient for thermal photons only. Recall that
\begin{eqnarray}
v_2 =  \int_0^{2 \pi} d \phi \cos \left( 2 \phi \right) \left( d^3 N/ d^2 p_T d y \right)/\int_0^{2 \pi} d \phi\, d^3 N/ d^2 p_T d y 
\end{eqnarray}
The net $v_2$ as a function of $p_T$ is shown, along with the individual QGP and HG coefficients, using averaged initial conditions (AIC, as discussed later). The right panel shows the effect of a finite shear viscosity coefficient on these quantities, after the 3+1 dimensional evolution. The influence of viscosity on this observable is remarkable, but it unfortunately acts to make a small signal even smaller, at least for conditions prevalent at RHIC. Thus, as is the case for hadrons, the momentum asymmetry of real photons is reduced by the shear viscous effects. 
\begin{figure}[h]
\begin{center}
{\includegraphics[width=7cm]{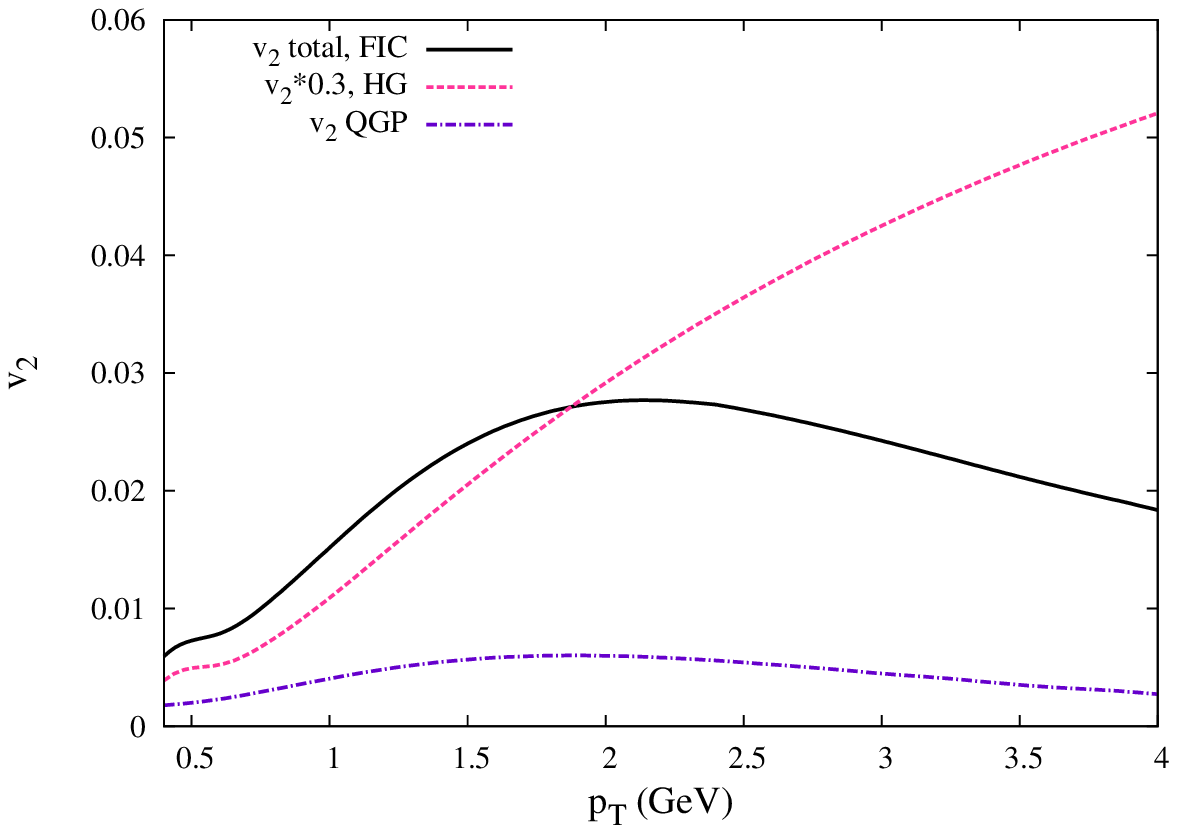}\hspace*{1cm}
\includegraphics[width=7cm]{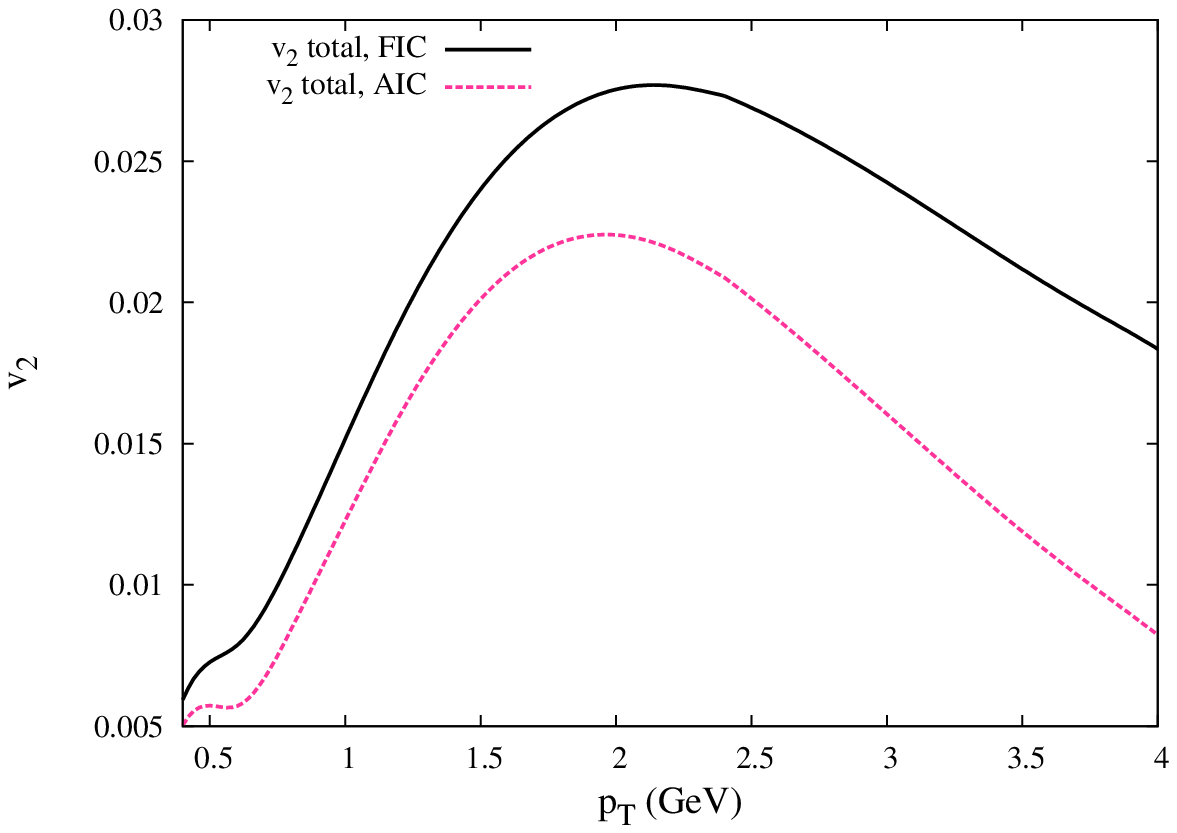}}
\caption{\sl  Left panel: The different $v_2$ coefficients obtained with fluctuating initial conditions (FIC). Right panel: The effect of AICs and FICs on the net $v_2$.  }
\label{FIC}
\end{center}
\end{figure}

The results of the calculations reported on up to now were all done assuming smooth (or averaged) initial state conditions in coordinate space. However, a   shift in paradigm occurred when recent analyses linked non-smooth, or fluctuating, initial conditions with the appearance of non-vanishing odd components of the momentum azimuthal asymmetry \cite{Alver:2010gr}.  The photon spectrum has been studied with fluctuating initial conditions (FIC) \cite{Chatterjee:2011rg} and will not be shown here for the sake of brevity.  We rather show results for the $v_2$ of real photons, starting from FIC, in Figure \ref{FIC}. In this low centrality class, one observes that the  FICs raise the photon $v_2$, as they do for hadrons \cite{Schenke:2010rr}. 

In conclusion, the results presented here constitute the first photon yield and $v_2$ obtained in a realistic 3D viscous hydrodynamics approach.  Our follow-up work will include more calculation details and applications to the LHC. Clearly, the continuing analyses at RHIC, the advent of LHC photon data, and the new quantitative link to realistic viscous hydrodynamics investigated here will fuel theoretical investigations for some time to come.

This work was supported in part by the Natural Sciences and Engineering Research Council of Canada, in part by the Fonds Nature et Technologies of the Government of Qu\'ebec, and in part by a Laboratory Directed Research and Development (LDRD) grant from Brookhaven Science Associates. 
\section*{References}
\bibliographystyle{iopart-num}
\bibliography{Gale}

\end{document}